\documentclass[twocolumn,aps,prx,twoside,superscriptaddress,floatfix,longbibliography,nofootinbib]{revtex4-1}

\usepackage{dcolumn}
\usepackage{float}
\usepackage{bm}
\usepackage[T1]{fontenc}
\usepackage{subcaption}
\usepackage{amsmath}
\usepackage{graphicx} 
\usepackage{tabularx}
\usepackage{multirow}
\usepackage{rotating} 
\usepackage{makecell}
\usepackage{hyperref}
\hypersetup{colorlinks=true, linkcolor=blue, citecolor=blue, urlcolor=blue}
\usepackage{tikz}
\usepackage[normalem]{ulem}

\usepackage[utf8]{inputenc}
\usepackage{booktabs}

\usepackage{color}

\begin{document}
\title{History of Ferroelectrics -- A Crystallography Perspective} 

\author{Nicola A. Spaldin}
\affiliation{Materials Theory, ETH Zurich, Wolfgang-Pauli-Strasse 28, 8093 Z\"urich, Switzerland}

\author{Ram Seshadri}
\affiliation{Materials Department, University of California, Santa Barbara CA 93106, USA}

\date{\today}
\begin{abstract}
``Underlying the whole treatment is the assumption that the physical properties of a solid are closely related to its structure, and that the first step to understanding the physical properties is to understand the structure.''\\
\hspace*{2.98in} Helen D. Megaw\\
\hspace*{2.98in} Preface to {\it Ferroelectricity in Crystals}, \\
\hspace*{3.62in} Methuen \& Co Ltd, London, 1957. \\

\begin{center}
To appear in the IUCr Newsletter
\end{center}

\end{abstract}
\maketitle

\begin{figure*}
    \centering
    \includegraphics[width=0.9\linewidth, trim={0 0 0cm 0},clip]{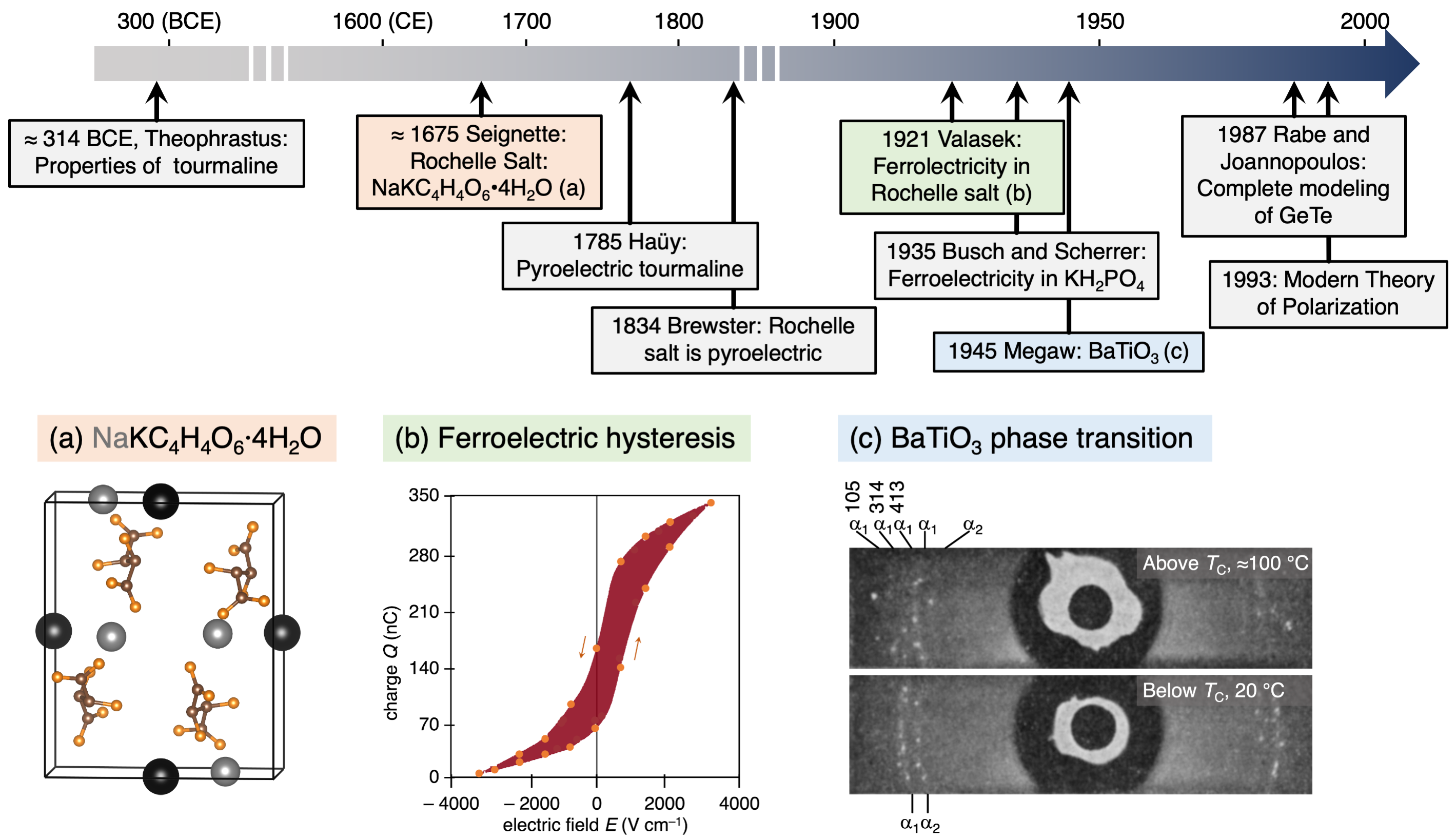}
    \caption{Upper panel: Timeline of some key events in the history of ferroelectrics, from 300 BCE to the present time. (a) Crystal structure of the paraelectric phase of Rochelle salt. Hydrogen atoms and water molecules are omitted for clarity. (b) The ferroelectric hysteresis loop of Rochelle salt recorded by Valasek in 1921 (data from reference \cite{Valasek:1921}). (c) A reproduction of the X-ray images recorded by Megaw with Cu $K\alpha_1$ and $K\alpha_2$ radiation \cite{Megaw:1946} of high-angle reflections in BaTiO$_3$ powders above and below the ferroelectric phase transition. \label{Timeline}}
\end{figure*}

2021 marks the 100th anniversary of the discovery of the phenomenon of ferroelectricity, defined to be a spontaneous electric polarization that is switchable by an applied electric field. This is a wonderful centennial for crystallographers to celebrate, because it is a property of crystals in which functionality and structure -- specifically the existence of a polar axis -- are intimately related. The key characteristic of a ferroelectric is a hysteresis loop of polarization versus electric field; we reproduce the data from the original discovery  by Joseph Valasek in Figure ~\ref{Timeline} \cite{Valasek:1921}. The similarity with the magnetization behavior of ferromagnetic iron in a magnetic field is clear -- just like the well-known hysteretic magnetic-field induced switching between the two stable states of opposite magnetization in a ferromagnet, a ferroelectric material has two stable states of opposite {\it electric} polarization associated with opposite relative displacements of its anions and cations -- and it's this analogy that gives ferroelectrics their name, in spite of the fact that most of them are not ferrous! We'll see in this article the key role played by crystallography in the evolution of ferroelectrics (a timeline of key discoveries is shown in Figure~\ref{Timeline}), starting from the identification of polar materials, through establishing the nature of ferroelectric phase transitions, to providing crucial information for the development of today's accepted theories of ferroelectric polarization.

We begin our story in 300 BC, when Theophrastus, the successor to Aristotle in his School of Philosophy, made an early mention of the electrostatic behavior of certain rocks in his treatise {\it On Stones}. Particularly intriguing is his report that ``lynx-urine stone'' (lyngourion) is better able to attract bits of straw, wood and metal if the stones are derived from wild, male lynx because of their superior diet and exercise regime \cite{Caley/Richards:1956}. The behavior he described is known today as {\it pyroelectricity}, and has the same  crystallographic requirement as that of ferroelectricity -- the presence of a polar axis -- although not all pyroelectrics are switchable by an applied field. For a discussion of the early history of the electrostatic behavior of pyroelectrics, we refer the reader to Ref.~\onlinecite{Lang:1974}). 

The first genuine crystallographic study of the pyroelectric effect occurred in 1785, when Abb\'e Ren\'e Just Ha\"uy, affectionately referred to as the father of crystallography~\cite{Walsh_book:1906}, turned his attention to the unusual pyroelectric behavior of the mineral tourmaline. The Abb\'e had by this time established basic laws of crystal form and symmetry through painstaking fragmentation of minerals to reveal their underlying crystallographic planes and angles. He therefore set about relating the pyroelectric behavior of tourmaline to its crystalline structure. First, he showed that the electricity in tourmaline was strongest at the poles of the crystal and became imperceptible at the middle, although when divided each new sub-crystal again had electricity at its poles. He also established that the existence of pyroelectricity correlates with an absence of symmetry in the crystal, which we now know provides the polar axis. In an early example of rational materials design, his finding led him to discover the effect in a number of other minerals. Charmingly, when his colleagues secured his release from imprisonment during the French Revolution, he was so absorbed in a crystallographic study that he refused to leave the prison until the next morning. We encourage PhD students to reflect on this on Friday afternoons, although we should perhaps disclose that the prisoners who remained were guillotined the following week.

Ha\"uy's flurry of discovery of new pyroelectrics was continued in particular by David Brewster, who, significantly for our discussion, identified pyroelectric behavior in sodium potassium tartrate tetrahydrate, more commonly known as Rochelle salt \cite{Brewster:1824} (Figure ~\ref{Timeline} a). First made in the 17th Century by adding sodium hydroxide to the tartar residue left in wine barrels, Rochelle salt was developed by the Seignette pharmaceutical dynasty of La Rochelle for use as a laxative. Thankfully, these authors can not comment on its effectiveness in this application.

In the century after pyroelectricity was identified in Rochelle Salt, a substantial body of experimental observations on this material accumulated. Some of the findings, particularly an anomalously large dielectric permittitivity and electrooptic Kerr effect \cite{Pockels_book:1894,Pockels_book:1906}, as well as hysteresis both in the piezoelectric response with pressure \cite{Cady:1918} and in the capacitance with charging direction \cite{Anderson:1918}, suggested an electrical resemblance to the ferromagnetic properties of iron \cite{Valasek:1971}. Valasek set about exploring the analogy, by measuring the charge on the plates of a Rochelle salt capacitor -- a proxy for the polarization of the material -- as a function of applied electric field. This simple experiment led him to discover the hysteresis of polarization with applied field reproduced in Figure~\ref{Timeline} \cite{Valasek:1921}, which is strongly reminiscent of the magnetization -- magnetic field behavior of a ferromagnet. Sadly, there was a resounding lack of interest given such a remarkable result, in part because Rochelle salt is unstable against dehydration and the results were not reproducible, and also because the behavior was not understood. Theoreticians were confounded by the crystal symmetry, which was incorrectly assigned on the basis of morphology to a non-centrosymmetric, but non-polar space group. With 112 atoms per unit cell a full x-ray crystal structure determination was attempted only twenty years later \cite{Beevers/Hughes:1941}. Even more unfortunately, the x-ray structure, being transparent to the symmetry-lowering water molecules, confirmed the incorrect assignment! (Not until decades later, when the first neutron diffraction structure on single crystals showed the positions of the polar water molecules that line up to provide the polarization, was the correct polar symmetry identified  \cite{Frazer/McKeown/Pepinsky:1954}.) And a peculiar behavior observed in only one badly behaved material offered only a rather rickety bandwagon for researchers to jump onto.

\begin{figure*}
    \begin{center}
    \includegraphics[width=0.75\linewidth, trim={0 0 0cm 0},clip]{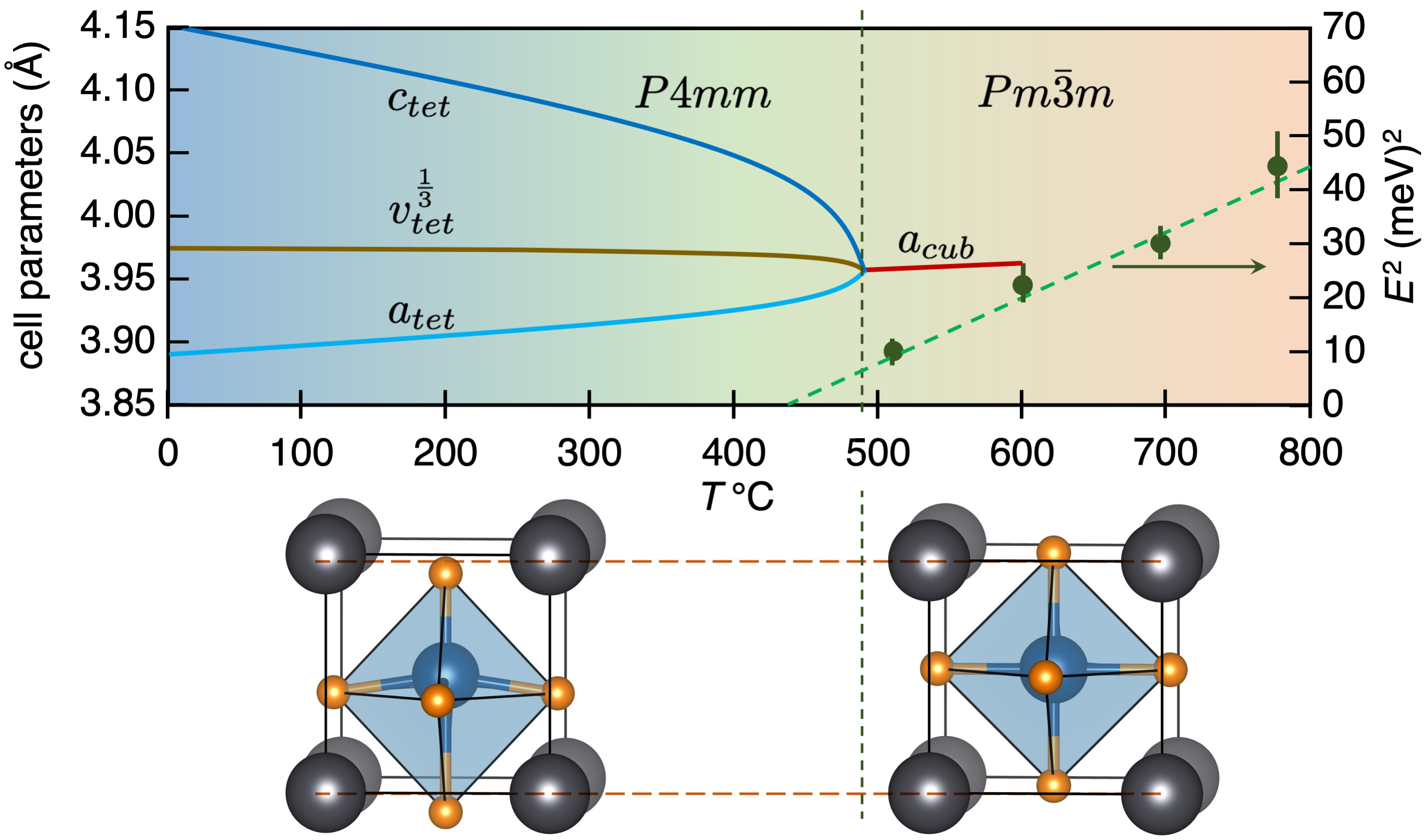}
    \end{center}
    \caption{Variation of the cell parameters of perovskite PbTiO$_3$ as a function of temperature across the phase transition (data adapted from reference \cite{Shirane/Hoshino/Suzuki:1950}). In the high-temperature cubic phase all three unit-cell lattice parameters have the same value, $a_{cub}$. In the low-temperature ferroelectric phase there is a tetragonal elongation with the lattice constant along the polar direction, $c_{tet}$, larger than those perpendicular to it, $a_{tet}$, and a small increase in volume, $V$. Data associated with the right ordinate depict the square of the soft mode energy as obtained from neutron scattering (data adapted from reference \cite{shirane1970soft}). The energy (or frequency) of this mode goes to zero near the transition temperature. The corresponding crystal structures of tetragonal ferroelectric PbTiO$_3$ at room temperature, and the cubic structure above the Curie temperature are depicted below. The Pb ions are shown in gray, the Ti in blue and the oxygen in orange. The loss of the dipole within the unit cell above the transition temperature is clearly seen from the oxygen octahedron around Ti becoming regular, as is the decrease in the $c$ lattice constant from the ferroelectric phase.  \label{PbTiO3}}
\end{figure*}

The lonely situation improved somewhat in 1935, when ferroelectricity was identified in a second, more  tractable material, potassium dihydrogen phosphate, KH$_2$PO$_4$ or KDP \cite{Busch/Scherrer:1935}. (Note that the authors called it a seignette-electric in homage to the La Rochelle Seignettes). In this case the crystal structure, except for the hydrogens, had already been determined using x-ray diffraction \cite{West:1930}, and later neutron studies found the hydrogen positions \cite{Bacon/Pease:1955} and showed that their ordering triggered the onset of ferroelectricity. 

The most significant breakthrough for the field, however, had to wait until the second world war, when materials with high dielectric constants were in demand as capacitor components in devices for military communications and detection. One such proprietary condenser sent to the UK Government from the United States mysteriously found its way into the Philips Materials Research Laboratory in Surrey, where it was pulled apart to yield a sample of barium titanate ceramic. Dielectric measurements quickly revealed a very high and temperature-dependent susceptibility, with a number of peaks suggestive of phase transitions. Powder X-ray diffraction was performed by Helen Megaw, who found a tetragonal structure at room temperature, and the ideal cubic perovksite structure above 120$^{\circ}$C \cite{Megaw:1945}, about which she writes ``I can still remember my thrill of pleasure in the dark room when I looked at the photo and saw that the lines were those of a simple cubic material!'' \cite{Megaw_Letter:1989}. Soon after, ferroelectric hysteresis was measured \cite{vonHippel_et_al:1946} and single crystals were produced \cite{Blattner_et_al:1947}. A robust material, with only five atoms per unit cell and ferroelectric properties, at room temperature, barium titanate BaTiO$_3$ and related compounds were almost too good to be true, allowing for detailed characterization as well as theoretical modeling. In particular, lead titanate PbTiO$_3$, which displays a relatively simple transition from the paraelectric cubic phase to a ferroelectric tetragonal phase just below 500\,$^\circ$C (Figure\,\ref{PbTiO3}), remains one of the highest-performing ferroelectric materials today. Collectively the ferroelectric titanates form the basis of an enormous industrial enterprise, with applications as diverse as transducers, capacitors, data storage and computer memories \cite{Scott:2007}.

Crystallographic measurements of the detailed atomic positions in BaTiO$_3$ were essential in developing the theory of ferroelectricity. In particular, the rather small displacement of the positively charged Ti ion relative to the negatively charged oxygen ions \cite{vonHippel:1950} showed that the polarization arose from rather subtle atomic displacements of the atoms, rather than the reorientation of discrete polar entities, and indicated the importance of the ionic polarizability in stabilizing the ferroelectric state \cite{Slater:1950}. This understanding enabled in turn the development of the so-called soft-mode theory of ferroelectricity \cite{Cochran:1959}, which recognizes that a crystal is only stable if displacements of its ions result in restoring forces that push them back to their starting positions. As the crystal approaches a phase transition, these forces become weaker, and the frequency of the corresponding lattice vibration lowers, or softens. In the ferroelectric titanates it's the polar phonon corresponding to the Ti - O relative displacement that softens and reaches zero frequency at the phase transition (Figure~\ref{PbTiO3}). Verifying this prediction was of course more of a source of excitement for spectroscopists than crystallographers \cite{Barker/Tinkham:1962,Cowley:1962}. With the advent of advanced electronic structure methods and improved computing capabilities, a full first-principles / statistical mechanical description of the ferroelectric phase transition became feasible and verified the soft-mode picture \cite{Rabe/Joannopoulos:1987}. Intriguingly, the most significant modern theoretical development -- the so-called modern theory of polarization -- shows that in fact only {\it changes} in polarization (for example between the cubic paraelectric and polar ferroelectric structures) are single-valued, and the general polarization in crystalline solids is in fact a {\it lattice} of values \cite{King-Smith/Vanderbilt:1993,Resta:1993,Spaldin:2012}. This elegantly closes the loop between experiment and theory, with the concepts of crystallography that were so essential in defining and identifying ferroelectricity, also proving relevant in its detailed quantum mechanical description.

So why did ferroelectricians have to wait until one hundred years ago for the emergence of their field? First, from a crystallographic standpoint, the fact that the polar state is usually formed on cooling from a non-polar parent phase leads to the formation of domains of different polar orientations. These polarization textures can be on the scale of tens of nanometers, rendering multi-domain crystals difficult to distinguish crystallographically from their centrosymmetric parent phases. In this context, powder diffraction can often be more revealing than single-crystal experiments. Indeed, the advent of the Rietveld method \cite{Rietveld:1969}, high-resolution X-ray powder diffraction at synchrotron sources \cite{Noheda_et_al:1999}, and the joint use of X-ray and neutron diffraction \cite{Corker_et_al:1997}, have collectively played an invaluable role in advancing the understanding of ferroic materials.

From a technical point of view, crystallographic measurements in finite electric fields are challenging, since (to quote Arthur Von Hippel) ``the crystallographer has thus to accept the unusual situation that the $c$-axis of the BaTiO$_3$ crystal can be turned around at will by the application of an electric field'' \cite{vonHippel:1950}. From a practical point of view, the engineering component in the definition of ferroelectricity -- that is the ability to switch with an electric field -- can not be assessed from knowledge of the crystal structure alone, and indeed before the availability of high quality thin-film samples, the voltages required for switching could be prohibitive.

While much is now understood about the phenomenology of ferroelectricity, there remain many open questions that are particularly relevant to crystallographers. Measurements that probe local structure, such as pair distribution function analysis \cite{Egami/Billinge:2003} and aberration-corrected electron microscopy \cite{SalmaniRezaie/Ahadi/Stemmer:2020,Bencan_et_al:2021} are starting to reveal previously hidden order such as polar nanodomains that are inaccessible to bulk techniques. Exotic non-uniform polarization arrangements reminiscent of skyrmions have been stabilized \cite{Das_et_al:2019} and incommensurate phases, in which the polar distortions do not match the crystallographic periodicity, identified \cite{Khalyavin_et_al:2020}. In the latter context, a general understanding of why most structural distortions are commensurate with the lattice is still lacking and a detailed knowledge of the crystallography of representative incommensurate phases would be invaluable. The excitement in the last decades in the area of multiferroic materials and phenomena, has been fuelled by intricate details of crystal structure revealing novel mechanisms for ferroelectricity, such as  the improper geometric behavior of YMnO$_3$ \cite{vanAken_et_al:2004} or the polar magnetic ordering in Cr$_2$BeO$_4$ \cite{Newnham_et_al:1978}. 

While many ferroelectrics in various crystal classes are now known, identifying new ones remains an active research area. Here crystallographic data bases provide invaluable input to modern high-throughput searches and machine learning studies. And finally, while ferroelectrics must be insulating so that an external electric field has a chance to switch the polarization, there is considerable current interest in polar metals, sometimes loosely referrred to as ferroelectric metals \cite{Anderson/Blount:1965}, because of their potentially exotic transport and superconductivity behaviors \cite{Enderlein_et_al:2020}. 

At its one hundredth birthday, the field of ferroelectrics is still thriving, with new phenomena and materials continuously being revealed and remaining to be discovered. Doubtless it will keep us busy for the next hundred years.

\bibliography{Nicola.bib}

\begin{thebibliography}{44}%
\makeatletter
\providecommand \@ifxundefined [1]{%
 \@ifx{#1\undefined}
}%
\providecommand \@ifnum [1]{%
 \ifnum #1\expandafter \@firstoftwo
 \else \expandafter \@secondoftwo
 \fi
}%
\providecommand \@ifx [1]{%
 \ifx #1\expandafter \@firstoftwo
 \else \expandafter \@secondoftwo
 \fi
}%
\providecommand \natexlab [1]{#1}%
\providecommand \enquote  [1]{``#1''}%
\providecommand \bibnamefont  [1]{#1}%
\providecommand \bibfnamefont [1]{#1}%
\providecommand \citenamefont [1]{#1}%
\providecommand \href@noop [0]{\@secondoftwo}%
\providecommand \href [0]{\begingroup \@sanitize@url \@href}%
\providecommand \@href[1]{\@@startlink{#1}\@@href}%
\providecommand \@@href[1]{\endgroup#1\@@endlink}%
\providecommand \@sanitize@url [0]{\catcode `\\12\catcode `\$12\catcode
  `\&12\catcode `\#12\catcode `\^12\catcode `\_12\catcode `\%12\relax}%
\providecommand \@@startlink[1]{}%
\providecommand \@@endlink[0]{}%
\providecommand \url  [0]{\begingroup\@sanitize@url \@url }%
\providecommand \@url [1]{\endgroup\@href {#1}{\urlprefix }}%
\providecommand \urlprefix  [0]{URL }%
\providecommand \Eprint [0]{\href }%
\providecommand \doibase [0]{http://dx.doi.org/}%
\providecommand \selectlanguage [0]{\@gobble}%
\providecommand \bibinfo  [0]{\@secondoftwo}%
\providecommand \bibfield  [0]{\@secondoftwo}%
\providecommand \translation [1]{[#1]}%
\providecommand \BibitemOpen [0]{}%
\providecommand \bibitemStop [0]{}%
\providecommand \bibitemNoStop [0]{.\EOS\space}%
\providecommand \EOS [0]{\spacefactor3000\relax}%
\providecommand \BibitemShut  [1]{\csname bibitem#1\endcsname}%
\let\auto@bib@innerbib\@empty
%</preamble>
\bibitem [{\citenamefont {Valasek}(1921)}]{Valasek:1921}%
  \BibitemOpen
  \bibfield  {author} {\bibinfo {author} {\bibfnamefont {J.}~\bibnamefont
  {Valasek}},\ }\bibfield  {title} {\enquote {\bibinfo {title} {Piezoelectric
  and allied phenomena in rochelle salt},}\ }\href@noop {} {\bibfield
  {journal} {\bibinfo  {journal} {Phys. Rev.}\ }\textbf {\bibinfo {volume}
  {17}},\ \bibinfo {pages} {475} (\bibinfo {year} {1921})}\BibitemShut
  {NoStop}%
\bibitem [{\citenamefont {Megaw}(1946)}]{Megaw:1946}%
  \BibitemOpen
  \bibfield  {author} {\bibinfo {author} {\bibfnamefont {H.~D.}\ \bibnamefont
  {Megaw}},\ }\bibfield  {title} {\enquote {\bibinfo {title} {Changes in
  polycrystalline barium-strontium titanate at its transition temperature},}\
  }\href@noop {} {\bibfield  {journal} {\bibinfo  {journal} {Nature}\ }\textbf
  {\bibinfo {volume} {157}},\ \bibinfo {pages} {20--21} (\bibinfo {year}
  {1946})}\BibitemShut {NoStop}%
\bibitem [{\citenamefont {Caley}\ and\ \citenamefont
  {Richards}(1956)}]{Caley/Richards:1956}%
  \BibitemOpen
  \bibfield  {author} {\bibinfo {author} {\bibfnamefont {E.~R.}\ \bibnamefont
  {Caley}}\ and\ \bibinfo {author} {\bibfnamefont {J.~F.~C.}\ \bibnamefont
  {Richards}},\ }\href@noop {} {\emph {\bibinfo {title} {Theophrastus On
  Stones}}}\ (\bibinfo  {publisher} {The Ohio State University},\ \bibinfo
  {year} {1956})\BibitemShut {NoStop}%
\bibitem [{\citenamefont {Lang}(1974)}]{Lang:1974}%
  \BibitemOpen
  \bibfield  {author} {\bibinfo {author} {\bibfnamefont {S.~B.}\ \bibnamefont
  {Lang}},\ }\bibfield  {title} {\enquote {\bibinfo {title} {{Pyroelectricity:
  A 2300-year history}},}\ }\href@noop {} {\bibfield  {journal} {\bibinfo
  {journal} {Ferroelectrics}\ }\textbf {\bibinfo {volume} {7}},\ \bibinfo
  {pages} {231--234} (\bibinfo {year} {1974})}\BibitemShut {NoStop}%
\bibitem [{\citenamefont {Walsh}(1906)}]{Walsh_book:1906}%
  \BibitemOpen
  \bibfield  {author} {\bibinfo {author} {\bibfnamefont {J.~J.}\ \bibnamefont
  {Walsh}},\ }\href@noop {} {\emph {\bibinfo {title} {{Catholic churchmen in
  science; sketches of the lives of Catholic ecclesiastics who were among the
  great founders in science}}}}\ (\bibinfo  {publisher} {Philadelphia: American
  Ecclesiastical Review},\ \bibinfo {year} {1906})\BibitemShut {NoStop}%
\bibitem [{\citenamefont {Brewster}(1824)}]{Brewster:1824}%
  \BibitemOpen
  \bibfield  {author} {\bibinfo {author} {\bibfnamefont {D.}~\bibnamefont
  {Brewster}},\ }\bibfield  {title} {\enquote {\bibinfo {title} {Observation of
  pyroelectricty in minerals},}\ }\href@noop {} {\bibfield  {journal} {\bibinfo
   {journal} {Edinburgh J. Sci.}\ }\textbf {\bibinfo {volume} {1}},\ \bibinfo
  {pages} {208--14} (\bibinfo {year} {1824})}\BibitemShut {NoStop}%
\bibitem [{\citenamefont {Pockels}(1894)}]{Pockels_book:1894}%
  \BibitemOpen
  \bibfield  {author} {\bibinfo {author} {\bibfnamefont {F.~C.~A.}\
  \bibnamefont {Pockels}},\ }\href@noop {} {\emph {\bibinfo {title} {{\"Uber
  den Einfluss des elektrostatischen Feldes auf das optische Verhalten
  piezoelektrischer Krystalle}}}}\ (\bibinfo  {publisher} {Dieterich,
  G\"ottingen},\ \bibinfo {year} {1894})\BibitemShut {NoStop}%
\bibitem [{\citenamefont {Pockels}(1906)}]{Pockels_book:1906}%
  \BibitemOpen
  \bibfield  {author} {\bibinfo {author} {\bibfnamefont {F.~C.~A.}\
  \bibnamefont {Pockels}},\ }\href@noop {} {\emph {\bibinfo {title} {{Lehrbuch
  der Kristalloptik}}}}\ (\bibinfo  {publisher} {Teubner, Leipzig},\ \bibinfo
  {year} {1906})\BibitemShut {NoStop}%
\bibitem [{\citenamefont {Cady}(1918)}]{Cady:1918}%
  \BibitemOpen
  \bibfield  {author} {\bibinfo {author} {\bibfnamefont {W.~G.}\ \bibnamefont
  {Cady}},\ }\href@noop {} {\enquote {\bibinfo {title} {{Rep. Nat. Research
  Council}},}\ } (\bibinfo {year} {1918})\BibitemShut {NoStop}%
\bibitem [{\citenamefont {Anderson}(1918)}]{Anderson:1918}%
  \BibitemOpen
  \bibfield  {author} {\bibinfo {author} {\bibfnamefont {J.~A.}\ \bibnamefont
  {Anderson}},\ }\href@noop {} {\enquote {\bibinfo {title} {{Rep. Nat. Research
  Council}},}\ } (\bibinfo {year} {1918})\BibitemShut {NoStop}%
\bibitem [{\citenamefont {Valasek}(1971)}]{Valasek:1971}%
  \BibitemOpen
  \bibfield  {author} {\bibinfo {author} {\bibfnamefont {J.}~\bibnamefont
  {Valasek}},\ }\bibfield  {title} {\enquote {\bibinfo {title} {The early
  history of ferroelectricity},}\ }\href@noop {} {\bibfield  {journal}
  {\bibinfo  {journal} {Ferroelectrics}\ }\textbf {\bibinfo {volume} {2}},\
  \bibinfo {pages} {239--244} (\bibinfo {year} {1971})}\BibitemShut {NoStop}%
\bibitem [{\citenamefont {Beevers}\ and\ \citenamefont
  {Hughes}(1941)}]{Beevers/Hughes:1941}%
  \BibitemOpen
  \bibfield  {author} {\bibinfo {author} {\bibfnamefont {C.~A.}\ \bibnamefont
  {Beevers}}\ and\ \bibinfo {author} {\bibfnamefont {W.}~\bibnamefont
  {Hughes}},\ }\href@noop {} {\bibfield  {journal} {\bibinfo  {journal} {Proc.
  Roy. Soc.}\ }\textbf {\bibinfo {volume} {177}},\ \bibinfo {pages} {251--259}
  (\bibinfo {year} {1941})}\BibitemShut {NoStop}%
\bibitem [{\citenamefont {Frazer}\ \emph {et~al.}(1954)\citenamefont {Frazer},
  \citenamefont {McKeown},\ and\ \citenamefont
  {Pepinsky}}]{Frazer/McKeown/Pepinsky:1954}%
  \BibitemOpen
  \bibfield  {author} {\bibinfo {author} {\bibfnamefont {B.~C.}\ \bibnamefont
  {Frazer}}, \bibinfo {author} {\bibfnamefont {M.}~\bibnamefont {McKeown}}, \
  and\ \bibinfo {author} {\bibfnamefont {R.}~\bibnamefont {Pepinsky}},\
  }\bibfield  {title} {\enquote {\bibinfo {title} {{Neutron diffraction studies
  of Rochelle-salt single crystals}},}\ }\href@noop {} {\bibfield  {journal}
  {\bibinfo  {journal} {Phys. Rev.}\ }\textbf {\bibinfo {volume} {94}},\
  \bibinfo {pages} {1435} (\bibinfo {year} {1954})}\BibitemShut {NoStop}%
\bibitem [{\citenamefont {Shirane}\ \emph {et~al.}(1950)\citenamefont
  {Shirane}, \citenamefont {Hoshino},\ and\ \citenamefont
  {Suzuki}}]{Shirane/Hoshino/Suzuki:1950}%
  \BibitemOpen
  \bibfield  {author} {\bibinfo {author} {\bibfnamefont {G.}~\bibnamefont
  {Shirane}}, \bibinfo {author} {\bibfnamefont {S.}~\bibnamefont {Hoshino}}, \
  and\ \bibinfo {author} {\bibfnamefont {K.}~\bibnamefont {Suzuki}},\
  }\bibfield  {title} {\enquote {\bibinfo {title} {X-ray study of the phase
  transition in lead titanate},}\ }\href@noop {} {\bibfield  {journal}
  {\bibinfo  {journal} {Phys. Rev.}\ }\textbf {\bibinfo {volume} {80}},\
  \bibinfo {pages} {1105} (\bibinfo {year} {1950})}\BibitemShut {NoStop}%
\bibitem [{\citenamefont {Shirane}\ \emph {et~al.}(1970)\citenamefont
  {Shirane}, \citenamefont {Axe}, \citenamefont {Harada},\ and\ \citenamefont
  {Remeika}}]{shirane1970soft}%
  \BibitemOpen
  \bibfield  {author} {\bibinfo {author} {\bibfnamefont {G}~\bibnamefont
  {Shirane}}, \bibinfo {author} {\bibfnamefont {JD}~\bibnamefont {Axe}},
  \bibinfo {author} {\bibfnamefont {J}~\bibnamefont {Harada}}, \ and\ \bibinfo
  {author} {\bibfnamefont {JP}~\bibnamefont {Remeika}},\ }\bibfield  {title}
  {\enquote {\bibinfo {title} {Soft ferroelectric modes in lead titanate},}\
  }\href@noop {} {\bibfield  {journal} {\bibinfo  {journal} {Phys. Rev. B}\
  }\textbf {\bibinfo {volume} {2}},\ \bibinfo {pages} {155} (\bibinfo {year}
  {1970})}\BibitemShut {NoStop}%
\bibitem [{\citenamefont {Busch}\ and\ \citenamefont
  {Scherrer}(1935)}]{Busch/Scherrer:1935}%
  \BibitemOpen
  \bibfield  {author} {\bibinfo {author} {\bibfnamefont {G.}~\bibnamefont
  {Busch}}\ and\ \bibinfo {author} {\bibfnamefont {P.}~\bibnamefont
  {Scherrer}},\ }\bibfield  {title} {\enquote {\bibinfo {title} {{A new
  seignette-electric substance}},}\ }\href@noop {} {\bibfield  {journal}
  {\bibinfo  {journal} {Naturwiss.}\ }\textbf {\bibinfo {volume} {23}},\
  \bibinfo {pages} {737} (\bibinfo {year} {1935})}\BibitemShut {NoStop}%
\bibitem [{\citenamefont {West}(1930)}]{West:1930}%
  \BibitemOpen
  \bibfield  {author} {\bibinfo {author} {\bibfnamefont {J.}~\bibnamefont
  {West}},\ }\bibfield  {title} {\enquote {\bibinfo {title} {{A quantitative
  X-ray analysis of the structure of potassium dihydrogen phosphate
  (KH$_2$PO$_4$)}},}\ }\href@noop {} {\bibfield  {journal} {\bibinfo  {journal}
  {Z. Kristallogr. Krist.}\ }\textbf {\bibinfo {volume} {74}},\ \bibinfo
  {pages} {306--332} (\bibinfo {year} {1930})}\BibitemShut {NoStop}%
\bibitem [{\citenamefont {Bacon}\ and\ \citenamefont
  {Pease}(1955)}]{Bacon/Pease:1955}%
  \BibitemOpen
  \bibfield  {author} {\bibinfo {author} {\bibfnamefont {G.~E.}\ \bibnamefont
  {Bacon}}\ and\ \bibinfo {author} {\bibfnamefont {R.~S.}\ \bibnamefont
  {Pease}},\ }\bibfield  {title} {\enquote {\bibinfo {title} {{A
  neutron-diffraction study of the ferroelectric transition of potassium
  dihydrogen phosphate}},}\ }\href@noop {} {\bibfield  {journal} {\bibinfo
  {journal} {Proc. Roy. Soc. A}\ }\textbf {\bibinfo {volume} {230}},\ \bibinfo
  {pages} {359--381} (\bibinfo {year} {1955})}\BibitemShut {NoStop}%
\bibitem [{\citenamefont {Megaw}(1945)}]{Megaw:1945}%
  \BibitemOpen
  \bibfield  {author} {\bibinfo {author} {\bibfnamefont {H.~D.}\ \bibnamefont
  {Megaw}},\ }\bibfield  {title} {\enquote {\bibinfo {title} {Crystal structure
  of barium titanate},}\ }\href@noop {} {\bibfield  {journal} {\bibinfo
  {journal} {Nature}\ }\textbf {\bibinfo {volume} {155}},\ \bibinfo {pages}
  {484} (\bibinfo {year} {1945})}\BibitemShut {NoStop}%
\bibitem [{\citenamefont {Megaw}(1989)}]{Megaw_Letter:1989}%
  \BibitemOpen
  \bibfield  {author} {\bibinfo {author} {\bibfnamefont {H.~D.}\ \bibnamefont
  {Megaw}},\ }\href@noop {} {\enquote {\bibinfo {title} {{Notes made by Helen
  D. Meghaw about some early work on ferroelectricity, together with some
  relevant pieces of autobiography}},}\ }\bibinfo {howpublished} {In a letter
  from Helen Meghaw to Bob Newnham, held in the Girton College, Cambridge
  Archives} (\bibinfo {year} {1989})\BibitemShut {NoStop}%
\bibitem [{\citenamefont {von Hippel}\ \emph {et~al.}(1946)\citenamefont {von
  Hippel}, \citenamefont {Breckenridge}, \citenamefont {Chesley},\ and\
  \citenamefont {Tisza}}]{vonHippel_et_al:1946}%
  \BibitemOpen
  \bibfield  {author} {\bibinfo {author} {\bibfnamefont {A.}~\bibnamefont {von
  Hippel}}, \bibinfo {author} {\bibfnamefont {R.~G.}\ \bibnamefont
  {Breckenridge}}, \bibinfo {author} {\bibfnamefont {F.~G.}\ \bibnamefont
  {Chesley}}, \ and\ \bibinfo {author} {\bibfnamefont {L.}~\bibnamefont
  {Tisza}},\ }\bibfield  {title} {\enquote {\bibinfo {title} {{High dielectric
  constant ceramics}},}\ }\href@noop {} {\bibfield  {journal} {\bibinfo
  {journal} {Ind. Eng. Chem.}\ }\textbf {\bibinfo {volume} {38}},\ \bibinfo
  {pages} {1097} (\bibinfo {year} {1946})}\BibitemShut {NoStop}%
\bibitem [{\citenamefont {Blattner}\ \emph {et~al.}(1947)\citenamefont
  {Blattner}, \citenamefont {Matthias}, \citenamefont {Merz},\ and\
  \citenamefont {Scherrer}}]{Blattner_et_al:1947}%
  \BibitemOpen
  \bibfield  {author} {\bibinfo {author} {\bibfnamefont {H.}~\bibnamefont
  {Blattner}}, \bibinfo {author} {\bibfnamefont {B.}~\bibnamefont {Matthias}},
  \bibinfo {author} {\bibfnamefont {W.}~\bibnamefont {Merz}}, \ and\ \bibinfo
  {author} {\bibfnamefont {P.}~\bibnamefont {Scherrer}},\ }\bibfield  {title}
  {\enquote {\bibinfo {title} {{Untersuchungen an
  Bariumtitanat-Einkristallen}},}\ }\href@noop {} {\bibfield  {journal}
  {\bibinfo  {journal} {Experientia}\ }\textbf {\bibinfo {volume} {3}},\
  \bibinfo {pages} {148} (\bibinfo {year} {1947})}\BibitemShut {NoStop}%
\bibitem [{\citenamefont {Scott}(2007)}]{Scott:2007}%
  \BibitemOpen
  \bibfield  {author} {\bibinfo {author} {\bibfnamefont {J.~F.}\ \bibnamefont
  {Scott}},\ }\bibfield  {title} {\enquote {\bibinfo {title} {Applications of
  modern ferroelectrics},}\ }\href@noop {} {\bibfield  {journal} {\bibinfo
  {journal} {Science}\ }\textbf {\bibinfo {volume} {315}},\ \bibinfo {pages}
  {954--959} (\bibinfo {year} {2007})}\BibitemShut {NoStop}%
\bibitem [{\citenamefont {von Hippel}(1950)}]{vonHippel:1950}%
  \BibitemOpen
  \bibfield  {author} {\bibinfo {author} {\bibfnamefont {A.}~\bibnamefont {von
  Hippel}},\ }\bibfield  {title} {\enquote {\bibinfo {title} {Ferroelectricity,
  domain structure, and phase transitions of barium titanate},}\ }\href@noop {}
  {\bibfield  {journal} {\bibinfo  {journal} {Rev. Mod. Phys.}\ }\textbf
  {\bibinfo {volume} {22}},\ \bibinfo {pages} {221--237} (\bibinfo {year}
  {1950})}\BibitemShut {NoStop}%
\bibitem [{\citenamefont {Slater}(1950)}]{Slater:1950}%
  \BibitemOpen
  \bibfield  {author} {\bibinfo {author} {\bibfnamefont {J.~C.}\ \bibnamefont
  {Slater}},\ }\bibfield  {title} {\enquote {\bibinfo {title} {{The Lorentz
  correction in barium titanate}},}\ }\href@noop {} {\bibfield  {journal}
  {\bibinfo  {journal} {Phys. Rev.}\ }\textbf {\bibinfo {volume} {78}},\
  \bibinfo {pages} {748} (\bibinfo {year} {1950})}\BibitemShut {NoStop}%
\bibitem [{\citenamefont {Cochran}(1959)}]{Cochran:1959}%
  \BibitemOpen
  \bibfield  {author} {\bibinfo {author} {\bibfnamefont {W.}~\bibnamefont
  {Cochran}},\ }\bibfield  {title} {\enquote {\bibinfo {title} {{Crystal
  stability and the theory of ferroelectricity}},}\ }\href@noop {} {\bibfield
  {journal} {\bibinfo  {journal} {Phys. Rev. Lett.}\ }\textbf {\bibinfo
  {volume} {3}},\ \bibinfo {pages} {412--414} (\bibinfo {year}
  {1959})}\BibitemShut {NoStop}%
\bibitem [{\citenamefont {Barker}\ and\ \citenamefont
  {Tinkham}(1962)}]{Barker/Tinkham:1962}%
  \BibitemOpen
  \bibfield  {author} {\bibinfo {author} {\bibfnamefont {A.~S.}\ \bibnamefont
  {Barker}}\ and\ \bibinfo {author} {\bibfnamefont {M.}~\bibnamefont
  {Tinkham}},\ }\bibfield  {title} {\enquote {\bibinfo {title} {{Far-infrared
  ferroelectric vibration mode in SrTiO$_3$}},}\ }\href@noop {} {\bibfield
  {journal} {\bibinfo  {journal} {Phys. Rev.}\ }\textbf {\bibinfo {volume}
  {125}},\ \bibinfo {pages} {1527} (\bibinfo {year} {1962})}\BibitemShut
  {NoStop}%
\bibitem [{\citenamefont {Cowley}(1962)}]{Cowley:1962}%
  \BibitemOpen
  \bibfield  {author} {\bibinfo {author} {\bibfnamefont {R.~A.}\ \bibnamefont
  {Cowley}},\ }\bibfield  {title} {\enquote {\bibinfo {title} {Temperature
  dependence of a transverse optic mode in strontium titanate},}\ }\href@noop
  {} {\bibfield  {journal} {\bibinfo  {journal} {Phys. Rev. Lett.}\ }\textbf
  {\bibinfo {volume} {9}},\ \bibinfo {pages} {159--161} (\bibinfo {year}
  {1962})}\BibitemShut {NoStop}%
\bibitem [{\citenamefont {Rabe}\ and\ \citenamefont
  {Joannopoulos}(1987)}]{Rabe/Joannopoulos:1987}%
  \BibitemOpen
  \bibfield  {author} {\bibinfo {author} {\bibfnamefont {K~M}\ \bibnamefont
  {Rabe}}\ and\ \bibinfo {author} {\bibfnamefont {J~D}\ \bibnamefont
  {Joannopoulos}},\ }\bibfield  {title} {\enquote {\bibinfo {title} {{Theory of
  the structural phase transition of GeTe}},}\ }\href@noop {} {\bibfield
  {journal} {\bibinfo  {journal} {Phys. Rev. B}\ }\textbf {\bibinfo {volume}
  {36}},\ \bibinfo {pages} {6631} (\bibinfo {year} {1987})}\BibitemShut
  {NoStop}%
\bibitem [{\citenamefont {King-Smith}\ and\ \citenamefont
  {Vanderbilt}(1993)}]{King-Smith/Vanderbilt:1993}%
  \BibitemOpen
  \bibfield  {author} {\bibinfo {author} {\bibfnamefont {R.~D.}\ \bibnamefont
  {King-Smith}}\ and\ \bibinfo {author} {\bibfnamefont {David}\ \bibnamefont
  {Vanderbilt}},\ }\bibfield  {title} {\enquote {\bibinfo {title} {Theory of
  polarization of crystalline solids},}\ }\href@noop {} {\bibfield  {journal}
  {\bibinfo  {journal} {Phys. Rev. B}\ }\textbf {\bibinfo {volume} {47}},\
  \bibinfo {pages} {R1651--R1654} (\bibinfo {year} {1993})}\BibitemShut
  {NoStop}%
\bibitem [{\citenamefont {Resta}(1993)}]{Resta:1993}%
  \BibitemOpen
  \bibfield  {author} {\bibinfo {author} {\bibfnamefont {R.}~\bibnamefont
  {Resta}},\ }\bibfield  {title} {\enquote {\bibinfo {title} {Macroscopic
  electric polarization as a geometric quantum phase},}\ }\href@noop {}
  {\bibfield  {journal} {\bibinfo  {journal} {Eur. Phys. Lett.}\ }\textbf
  {\bibinfo {volume} {22}},\ \bibinfo {pages} {133--138} (\bibinfo {year}
  {1993})}\BibitemShut {NoStop}%
\bibitem [{\citenamefont {Spaldin}(2012)}]{Spaldin:2012}%
  \BibitemOpen
  \bibfield  {author} {\bibinfo {author} {\bibfnamefont {N.~A.}\ \bibnamefont
  {Spaldin}},\ }\bibfield  {title} {\enquote {\bibinfo {title} {A beginner's
  guide to the modern theory of polarization},}\ }\href@noop {} {\bibfield
  {journal} {\bibinfo  {journal} {J.\ Sol.\ Stat.\ Chem.}\ }\textbf {\bibinfo
  {volume} {195}},\ \bibinfo {pages} {2} (\bibinfo {year} {2012})}\BibitemShut
  {NoStop}%
\bibitem [{\citenamefont {Rietveld}(1969)}]{Rietveld:1969}%
  \BibitemOpen
  \bibfield  {author} {\bibinfo {author} {\bibfnamefont {H.~M.}\ \bibnamefont
  {Rietveld}},\ }\bibfield  {title} {\enquote {\bibinfo {title} {A profile
  refinement method for nuclear and magnetic structures},}\ }\href@noop {}
  {\bibfield  {journal} {\bibinfo  {journal} {J. Appl. Crystallogr.}\ }\textbf
  {\bibinfo {volume} {2}},\ \bibinfo {pages} {65--71} (\bibinfo {year}
  {1969})}\BibitemShut {NoStop}%
\bibitem [{\citenamefont {Noheda}\ \emph {et~al.}(1999)\citenamefont {Noheda},
  \citenamefont {Cox}, \citenamefont {Shirane}, \citenamefont {Gonzalo},
  \citenamefont {Cross},\ and\ \citenamefont {Park}}]{Noheda_et_al:1999}%
  \BibitemOpen
  \bibfield  {author} {\bibinfo {author} {\bibfnamefont {B.}~\bibnamefont
  {Noheda}}, \bibinfo {author} {\bibfnamefont {D.~E.}\ \bibnamefont {Cox}},
  \bibinfo {author} {\bibfnamefont {G.}~\bibnamefont {Shirane}}, \bibinfo
  {author} {\bibfnamefont {J.~A.}\ \bibnamefont {Gonzalo}}, \bibinfo {author}
  {\bibfnamefont {L.~E.}\ \bibnamefont {Cross}}, \ and\ \bibinfo {author}
  {\bibfnamefont {S.~E.}\ \bibnamefont {Park}},\ }\bibfield  {title} {\enquote
  {\bibinfo {title} {{A monoclinic ferroelectric phase in the
  Pb(Zr$_{1-x}$Ti$_x$)O$_3$ solid solution}},}\ }\href@noop {} {\bibfield
  {journal} {\bibinfo  {journal} {Appl. Phys. Lett.}\ }\textbf {\bibinfo
  {volume} {74}},\ \bibinfo {pages} {2059--2061} (\bibinfo {year}
  {1999})}\BibitemShut {NoStop}%
\bibitem [{\citenamefont {Corker}\ \emph {et~al.}(1997)\citenamefont {Corker},
  \citenamefont {Glazer}, \citenamefont {Dec}, \citenamefont {Roleder},\ and\
  \citenamefont {Whatmore}}]{Corker_et_al:1997}%
  \BibitemOpen
  \bibfield  {author} {\bibinfo {author} {\bibfnamefont {D.~L.}\ \bibnamefont
  {Corker}}, \bibinfo {author} {\bibfnamefont {A.~M.}\ \bibnamefont {Glazer}},
  \bibinfo {author} {\bibfnamefont {J.}~\bibnamefont {Dec}}, \bibinfo {author}
  {\bibfnamefont {K.}~\bibnamefont {Roleder}}, \ and\ \bibinfo {author}
  {\bibfnamefont {R.~W.}\ \bibnamefont {Whatmore}},\ }\bibfield  {title}
  {\enquote {\bibinfo {title} {{A re-investigation of the crystal structure of
  the perovskite PbZrO$_3$ by X-ray and neutron diffraction}},}\ }\href@noop {}
  {\bibfield  {journal} {\bibinfo  {journal} {Acta Crystallogr., Sect. B:
  Struct. Sci.}\ }\textbf {\bibinfo {volume} {53}},\ \bibinfo {pages}
  {135--142} (\bibinfo {year} {1997})}\BibitemShut {NoStop}%
\bibitem [{\citenamefont {Egami}\ and\ \citenamefont
  {Billinge}(2003)}]{Egami/Billinge:2003}%
  \BibitemOpen
  \bibfield  {author} {\bibinfo {author} {\bibfnamefont {T.}~\bibnamefont
  {Egami}}\ and\ \bibinfo {author} {\bibfnamefont {S.~J.~L.}\ \bibnamefont
  {Billinge}},\ }\href@noop {} {\emph {\bibinfo {title} {{Underneath the Bragg
  peaks: structural analysis of complex materials}}}}\ (\bibinfo  {publisher}
  {Elsevier},\ \bibinfo {year} {2003})\BibitemShut {NoStop}%
\bibitem [{\citenamefont {Salmani-Rezaie}\ \emph {et~al.}(2020)\citenamefont
  {Salmani-Rezaie}, \citenamefont {Ahadi},\ and\ \citenamefont
  {Stemmer}}]{SalmaniRezaie/Ahadi/Stemmer:2020}%
  \BibitemOpen
  \bibfield  {author} {\bibinfo {author} {\bibfnamefont {S.}~\bibnamefont
  {Salmani-Rezaie}}, \bibinfo {author} {\bibfnamefont {K.}~\bibnamefont
  {Ahadi}}, \ and\ \bibinfo {author} {\bibfnamefont {S.}~\bibnamefont
  {Stemmer}},\ }\bibfield  {title} {\enquote {\bibinfo {title} {{Polar
  nanodomains in a ferroelectric superconductor}},}\ }\href@noop {} {\bibfield
  {journal} {\bibinfo  {journal} {Nano Lett.}\ }\textbf {\bibinfo {volume}
  {20}},\ \bibinfo {pages} {6542--6547} (\bibinfo {year} {2020})}\BibitemShut
  {NoStop}%
\bibitem [{\citenamefont {Bencan}\ \emph {et~al.}(2021)\citenamefont {Bencan},
  \citenamefont {Oveisi}, \citenamefont {Hashemizadeh}, \citenamefont
  {Veerapandiyan}, \citenamefont {Hoshina}, \citenamefont {Rojac},
  \citenamefont {Deluca}, \citenamefont {Drazic},\ and\ \citenamefont
  {Damjanovic}}]{Bencan_et_al:2021}%
  \BibitemOpen
  \bibfield  {author} {\bibinfo {author} {\bibfnamefont {A.}~\bibnamefont
  {Bencan}}, \bibinfo {author} {\bibfnamefont {E.}~\bibnamefont {Oveisi}},
  \bibinfo {author} {\bibfnamefont {S.}~\bibnamefont {Hashemizadeh}}, \bibinfo
  {author} {\bibfnamefont {V.~K}\ \bibnamefont {Veerapandiyan}}, \bibinfo
  {author} {\bibfnamefont {T.}~\bibnamefont {Hoshina}}, \bibinfo {author}
  {\bibfnamefont {T.}~\bibnamefont {Rojac}}, \bibinfo {author} {\bibfnamefont
  {M.}~\bibnamefont {Deluca}}, \bibinfo {author} {\bibfnamefont
  {G.}~\bibnamefont {Drazic}}, \ and\ \bibinfo {author} {\bibfnamefont
  {D.}~\bibnamefont {Damjanovic}},\ }\bibfield  {title} {\enquote {\bibinfo
  {title} {Atomic scale symmetry and polar nanoclusters in the paraelectric
  phase of ferroelectric materials},}\ }\href@noop {} {\bibfield  {journal}
  {\bibinfo  {journal} {Nat. Commun.}\ }\textbf {\bibinfo {volume} {12}},\
  \bibinfo {pages} {1--9} (\bibinfo {year} {2021})}\BibitemShut {NoStop}%
\bibitem [{\citenamefont {Das}\ \emph {et~al.}(2019)\citenamefont {Das},
  \citenamefont {Tang}, \citenamefont {Hong}, \citenamefont {Gon{\c c}alves},
  \citenamefont {McCarter}, \citenamefont {Klewe}, \citenamefont {Nguyen},
  \citenamefont {G{\'o}mez-Ortiz}, \citenamefont {Shafer}, \citenamefont
  {Arenholz}, \citenamefont {Stoica}, \citenamefont {Hsu}, \citenamefont
  {Wang}, \citenamefont {Ophus}, \citenamefont {Liu}, \citenamefont {Nelson},
  \citenamefont {Saremi}, \citenamefont {Prasad}, \citenamefont {Mei},
  \citenamefont {Schlom}, \citenamefont {{\'I}{\~n}iguez}, \citenamefont
  {Garcia-Fernandez}, \citenamefont {Muller}, \citenamefont {Chen},
  \citenamefont {Junquera}, \citenamefont {Martin},\ and\ \citenamefont
  {Ramesh}}]{Das_et_al:2019}%
  \BibitemOpen
  \bibfield  {author} {\bibinfo {author} {\bibfnamefont {S.}~\bibnamefont
  {Das}}, \bibinfo {author} {\bibfnamefont {Y.~L.}\ \bibnamefont {Tang}},
  \bibinfo {author} {\bibfnamefont {Z.}~\bibnamefont {Hong}}, \bibinfo {author}
  {\bibfnamefont {M.~A.~P.}\ \bibnamefont {Gon{\c c}alves}}, \bibinfo {author}
  {\bibfnamefont {M.~R.}\ \bibnamefont {McCarter}}, \bibinfo {author}
  {\bibfnamefont {C.}~\bibnamefont {Klewe}}, \bibinfo {author} {\bibfnamefont
  {K.~X.}\ \bibnamefont {Nguyen}}, \bibinfo {author} {\bibfnamefont
  {F.}~\bibnamefont {G{\'o}mez-Ortiz}}, \bibinfo {author} {\bibfnamefont
  {P.}~\bibnamefont {Shafer}}, \bibinfo {author} {\bibfnamefont
  {E.}~\bibnamefont {Arenholz}}, \bibinfo {author} {\bibfnamefont {V.~A.}\
  \bibnamefont {Stoica}}, \bibinfo {author} {\bibfnamefont {S.~L.}\
  \bibnamefont {Hsu}}, \bibinfo {author} {\bibfnamefont {B.}~\bibnamefont
  {Wang}}, \bibinfo {author} {\bibfnamefont {C.}~\bibnamefont {Ophus}},
  \bibinfo {author} {\bibfnamefont {J.~F.}\ \bibnamefont {Liu}}, \bibinfo
  {author} {\bibfnamefont {C.~T.}\ \bibnamefont {Nelson}}, \bibinfo {author}
  {\bibfnamefont {S.}~\bibnamefont {Saremi}}, \bibinfo {author} {\bibfnamefont
  {B.}~\bibnamefont {Prasad}}, \bibinfo {author} {\bibfnamefont {A.~B.}\
  \bibnamefont {Mei}}, \bibinfo {author} {\bibfnamefont {D.~G.}\ \bibnamefont
  {Schlom}}, \bibinfo {author} {\bibfnamefont {J.}~\bibnamefont
  {{\'I}{\~n}iguez}}, \bibinfo {author} {\bibfnamefont {P.}~\bibnamefont
  {Garcia-Fernandez}}, \bibinfo {author} {\bibfnamefont {D.~A.}\ \bibnamefont
  {Muller}}, \bibinfo {author} {\bibfnamefont {L.~Q.}\ \bibnamefont {Chen}},
  \bibinfo {author} {\bibfnamefont {J.}~\bibnamefont {Junquera}}, \bibinfo
  {author} {\bibfnamefont {L.~W.}\ \bibnamefont {Martin}}, \ and\ \bibinfo
  {author} {\bibfnamefont {R.}~\bibnamefont {Ramesh}},\ }\bibfield  {title}
  {\enquote {\bibinfo {title} {{Observation of room-temperature polar
  skyrmions}},}\ }\href@noop {} {\bibfield  {journal} {\bibinfo  {journal}
  {Nature}\ }\textbf {\bibinfo {volume} {568}},\ \bibinfo {pages} {368--372}
  (\bibinfo {year} {2019})}\BibitemShut {NoStop}%
\bibitem [{\citenamefont {Khalyavin}\ \emph {et~al.}(2020)\citenamefont
  {Khalyavin}, \citenamefont {Johnson}, \citenamefont {Orlandi}, \citenamefont
  {Radaelli}, \citenamefont {Manuel},\ and\ \citenamefont
  {Belik}}]{Khalyavin_et_al:2020}%
  \BibitemOpen
  \bibfield  {author} {\bibinfo {author} {\bibfnamefont {D.~D.}\ \bibnamefont
  {Khalyavin}}, \bibinfo {author} {\bibfnamefont {R.~D.}\ \bibnamefont
  {Johnson}}, \bibinfo {author} {\bibfnamefont {F.}~\bibnamefont {Orlandi}},
  \bibinfo {author} {\bibfnamefont {P.~aG.}\ \bibnamefont {Radaelli}}, \bibinfo
  {author} {\bibfnamefont {P.}~\bibnamefont {Manuel}}, \ and\ \bibinfo {author}
  {\bibfnamefont {A.~A.}\ \bibnamefont {Belik}},\ }\bibfield  {title} {\enquote
  {\bibinfo {title} {{Emergent helical texture of electric dipoles}},}\
  }\href@noop {} {\bibfield  {journal} {\bibinfo  {journal} {Science}\ }\textbf
  {\bibinfo {volume} {369}},\ \bibinfo {pages} {680--684} (\bibinfo {year}
  {2020})}\BibitemShut {NoStop}%
\bibitem [{\citenamefont {van Aken}\ \emph {et~al.}(2004)\citenamefont {van
  Aken}, \citenamefont {Palstra}, \citenamefont {Filippetti},\ and\
  \citenamefont {Spaldin}}]{vanAken_et_al:2004}%
  \BibitemOpen
  \bibfield  {author} {\bibinfo {author} {\bibfnamefont {B.~B.}\ \bibnamefont
  {van Aken}}, \bibinfo {author} {\bibfnamefont {T.~T.~M.}\ \bibnamefont
  {Palstra}}, \bibinfo {author} {\bibfnamefont {A.}~\bibnamefont {Filippetti}},
  \ and\ \bibinfo {author} {\bibfnamefont {N.~A.}\ \bibnamefont {Spaldin}},\
  }\bibfield  {title} {\enquote {\bibinfo {title} {The origin of
  ferroelectricity in magnetoelectric {Y}{M}n{O}$_3$},}\ }\href@noop {}
  {\bibfield  {journal} {\bibinfo  {journal} {Nat. Mater.}\ }\textbf {\bibinfo
  {volume} {3}},\ \bibinfo {pages} {164--170} (\bibinfo {year}
  {2004})}\BibitemShut {NoStop}%
\bibitem [{\citenamefont {Newnham}\ \emph {et~al.}(1978)\citenamefont
  {Newnham}, \citenamefont {Kramer}, \citenamefont {Schulze},\ and\
  \citenamefont {Cross}}]{Newnham_et_al:1978}%
  \BibitemOpen
  \bibfield  {author} {\bibinfo {author} {\bibfnamefont {R.~E.}\ \bibnamefont
  {Newnham}}, \bibinfo {author} {\bibfnamefont {J.~J.}\ \bibnamefont {Kramer}},
  \bibinfo {author} {\bibfnamefont {W.~E.}\ \bibnamefont {Schulze}}, \ and\
  \bibinfo {author} {\bibfnamefont {L.~E.}\ \bibnamefont {Cross}},\ }\bibfield
  {title} {\enquote {\bibinfo {title} {Magnetoferroelectricity in
  {Cr$_2$BeO$_4$}},}\ }\href@noop {} {\bibfield  {journal} {\bibinfo  {journal}
  {J. Appl. Phys.}\ }\textbf {\bibinfo {volume} {49}},\ \bibinfo {pages}
  {6088--6091} (\bibinfo {year} {1978})}\BibitemShut {NoStop}%
\bibitem [{\citenamefont {Anderson}\ and\ \citenamefont
  {Blount}(1965)}]{Anderson/Blount:1965}%
  \BibitemOpen
  \bibfield  {author} {\bibinfo {author} {\bibfnamefont {P.~W.}\ \bibnamefont
  {Anderson}}\ and\ \bibinfo {author} {\bibfnamefont {E.~I.}\ \bibnamefont
  {Blount}},\ }\bibfield  {title} {\enquote {\bibinfo {title} {Symmetry
  considerations on martensitic transformations: ``ferroelectric'' metals?}}\
  }\href@noop {} {\bibfield  {journal} {\bibinfo  {journal} {Phys. Rev. Lett.}\
  }\textbf {\bibinfo {volume} {14}},\ \bibinfo {pages} {217} (\bibinfo {year}
  {1965})}\BibitemShut {NoStop}%
\bibitem [{\citenamefont {Enderlein}\ \emph {et~al.}(2020)\citenamefont
  {Enderlein}, \citenamefont {de~Oliveira}, \citenamefont {Tompsett},
  \citenamefont {Saitovitch}, \citenamefont {Saxena}, \citenamefont
  {Lonzarich},\ and\ \citenamefont {Rowley}}]{Enderlein_et_al:2020}%
  \BibitemOpen
  \bibfield  {author} {\bibinfo {author} {\bibfnamefont {C.}~\bibnamefont
  {Enderlein}}, \bibinfo {author} {\bibfnamefont {J.~F.}\ \bibnamefont
  {de~Oliveira}}, \bibinfo {author} {\bibfnamefont {D.~A.}\ \bibnamefont
  {Tompsett}}, \bibinfo {author} {\bibfnamefont {E.~B.}\ \bibnamefont
  {Saitovitch}}, \bibinfo {author} {\bibfnamefont {S.~S.}\ \bibnamefont
  {Saxena}}, \bibinfo {author} {\bibfnamefont {G.~G.}\ \bibnamefont
  {Lonzarich}}, \ and\ \bibinfo {author} {\bibfnamefont {S.~E.}\ \bibnamefont
  {Rowley}},\ }\bibfield  {title} {\enquote {\bibinfo {title}
  {Superconductivity mediated by polar modes in ferroelectric metals},}\
  }\href@noop {} {\bibfield  {journal} {\bibinfo  {journal} {Nat. Commun.}\
  }\textbf {\bibinfo {volume} {11}},\ \bibinfo {pages} {1--10} (\bibinfo {year}
  {2020})}\BibitemShut {NoStop}%
\end{thebibliography}%
\end{document}